\documentclass[sigconf, nonacm]{acmart}
\usepackage{balance} 
\usepackage{graphicx}
\usepackage{textcomp}
\usepackage{xcolor}
\usepackage{booktabs}
\usepackage{enumitem}
\usepackage{hyperref}
\usepackage{totpages}
\usepackage{subcaption}

\AtBeginDocument{%
  \providecommand\BibTeX{{%
    \normalfont B\kern-0.5em{\scshape i\kern-0.25em b}\kern-0.8em\TeX}}}

\begin{document}

\graphicspath{ {./img/} }

\newlist{questions}{enumerate}{2}
\setlist[questions,1]{label=\textbf{RQ\arabic*.},ref=RQ\arabic*}
\setlist[questions,2]{label=(\alph*),ref=\thequestionsi(\alph*)}

\title[Survival Analysis of Open Source Projects]{Two Approaches to Survival Analysis of Open Source Python Projects}

\author{Derek Robinson}
\affiliation{%
    \institution{University of Victoria}
    \department{Computer Science}
    \city{Victoria}
    \country(Canada)
}
\email{drobinson@uvic.ca}

\author{Keanelek Enns}
\affiliation{%
    \institution{University of Victoria}
    \department{Computer Science}
    \city{Victoria}
    \country(Canada)
}
\email{keanelekenns@uvic.ca}

\author{Neha Koulecar}
\affiliation{%
    \institution{University of Victoria}
    \department{Computer Science}
    \city{Victoria}
    \country(Canada)
}
\email{nehakoulecar@uvic.ca}

\author{Manish Sihag}
\affiliation{%
    \institution{University of Victoria}
    \department{Computer Science}
    \city{Victoria}
    \country(Canada)
}
\email{manishsihag@uvic.ca}
\renewcommand{\shortauthors}{D. Robinson, K. Enns, N. Koulecar, M. Sihag}

\begin{abstract}
    A recent study applied frequentist survival analysis methods to a subset of the Software Heritage Graph and determined which attributes of an OSS project contribute to its health. 
    This paper serves as an exact replication of that study. 
    In addition, Bayesian survival analysis methods were applied to the same dataset, and an additional project attribute was studied to serve as a conceptual replication.
    Both analyses focus on the effects of certain attributes on the survival of open-source software projects as measured by their revision activity.
    Methods such as the Kaplan-Meier estimator, Cox Proportional-Hazards model, and the visualization of posterior survival functions were used for each of the project attributes.
    The results show that projects which publish major releases, have repositories on multiple hosting services, possess a large team of developers, and make frequent revisions have a higher likelihood of survival in the long run.
    The findings were similar to the original study; however, a deeper look revealed quantitative inconsistencies.
\end{abstract}

\ccsdesc[500]{Software and its engineering~Open source model}
\ccsdesc[300]{Information systems~Data mining}

\keywords{data science, survival analysis, open source, python, Kaplan Meier, Cox Proportional-Hazards model, Bayesian analysis, frequentist}

\settopmatter{printfolios=true} 

\maketitle

\section{Introduction} \label{intro}

The developers of open-source software (OSS) projects are often part of decentralized and geographically distributed teams of volunteers.
As these developers volunteer their free time to build such OSS projects, they likely want to be confident that the projects they work on will not become inactive.
Suppose OSS developers are aware of key attributes that are associated with long-lasting projects.
In that case, they can make informed assessments of a given project before devoting their time to it, or they can strive to make their own projects exhibit those attributes.
Understanding which attributes of an OSS project lead to its longevity motivated Ali \emph{et al.} to apply survival analysis techniques commonly found in biostatistics to study the probability of survival for popular OSS Python projects \cite{ali2020cheating}.
Ali \emph{et al.} (referred to as the original authors from here on) specifically studied the effect of the following attributes on the survival of OSS Python projects: publishing major releases, the use of multiple hosting services, the type of hosting service, and the size of the volunteer team.

Survival analysis is a set of methods used to determine how long an entity will live (or the time to a given event of interest, such as death) and is often used in the medical field.
For example, survival analysis can determine the probability of a patient surviving past a certain time when given a treatment.
However, death is not as well defined for a software project as it is for a living organism.
A project may not receive revisions for an extended period only to be returned to at a later date, or perhaps a project no longer receives revisions at all, but the community that uses it continues to be active.
Samoladas \emph{et al.} considered a project inactive if it received less than two revisions a month; two months of inactivity led to it being considered abandoned or dead \cite{samoladas2010survival}.
Evangelopoulos \emph{et al.}~\cite{evangelopoulos} and the original authors~\cite{ali2020cheating} considered a project dead once there were no revisions at all.
The latter definition of project abandonment or death is used in this study to measure the duration of a project or its survival.

The original authors use a frequentist approach to survival analysis utilizing such methods as the Kaplan-Meier (K-M) survival estimator~\cite{kaplan1958nonparametric} and the Cox Proportional-Hazards model~\cite{cox1972regression}.
Though frequentist approaches are considered to be unbiased, minimal in variance, efficient, and generally sufficient, some consider them to lack robustness \cite{renganathan2016overview}. Another approach to survival analysis, Bayesian analysis, is considered to generate more robust models that perform well under the presence of new data being introduced and are generally easier to interpret results from \cite{renganathan2016overview}.

The authors of this paper resonate with the motivation of the original authors. 
This paper serves as an exact replication \cite{shull2008role} of their paper \cite{ali2020cheating} (referred to as the original paper from here on) and seeks to reproduce their analyses. 
This replication also provides artifacts so that others may see how the study was conducted and reproduce it with ease.
In addition to studying the same attributes as the original authors, the revision frequency of a given project was also studied.
Furthermore, this paper analyzes the same data set using a Bayesian approach to survival analysis as outlined by Kelter \cite{kelter2020bayesian} and seeks to compare the results of the frequentist and Bayesian approaches in the same domain.
The additional revision frequency analysis and Bayesian analysis serve as a conceptual replication \cite{shull2008role} of the original study \cite{ali2020cheating}.
Thus, the research questions this paper answers are as follows:

\begin{questions}
    \item How do major releases, the use of multiple hosting services, the type of hosting service, and the size of the volunteer team affect the probability of survival of an OSS Python project?
    \item How does the revision frequency of an OSS Python project affect the probability of its survival?
    \item How do the findings of frequentist survival analysis differ from those of Bayesian survival analysis?
\end{questions}

The remainder of the paper is structured as follows. 
Section \ref{related} outlines other research which has utilized survival analysis to study OSS and other work which has studied attributes similar to those studied in this paper.
Section \ref{data} describes the source of the data set, the data set itself, and the required preparation in order to perform survival analysis.
Section \ref{methods} covers the methods used for the replication, Bayesian analysis, and the additional attribute analysis.
Section \ref{results} shows the results for each analysis.
Section \ref{discussion} discusses the results, implications, and limitations of the analyses, and gives suggestions for future work.
The final section concludes by summarizing the purpose and findings of this paper.

\section{Related Work} \label{related}
Several other researchers have employed survival analysis to study the health of OSS projects. 
For example, Samoladas \emph{et al.} studied the effect of application domain and developer count on OSS project health, which was measured using project duration \cite{samoladas2010survival}. 
They found that applications within the domain of \emph{games and entertainment} and \emph{security} had the lowest probability of survival. 
Additionally, they found that for each new developer introduced to a project, the projects survivability increased by 15.8\%.
On the topic of developers, several studies have used survival analysis to study developer disengagement from OSS projects \cite{miller2019people,lin2017developer,ortega2009survival}. 
Miller \emph{et al.} made use of a survey and survival analyses, to determine the causes behind why a developer might stop contributing to an OSS project \cite{miller2019people}. 
Their analysis revealed that developers have a higher probability of project disengagement when going through job transitions and when working longer hours. 
Lin \emph{et al.} determined that developers who maintained files created by both themselves and others have a higher survival probability than developers who only maintain their own files or only maintain others files \cite{lin2017developer}. 
Additionally, Lin \emph{et al.} found that developers who maintained files and developers who mainly wrote code had a higher survival probability than those who solely created files and those whose main focus was writing documentation \cite{lin2017developer}. 
Ortega and Izquierdo-Cortazar analyzed the survival of OSS committers and Wikipedia editors and found that OSS committers have higher mean survival times than Wikipedia editors \cite{ortega2009survival}. 

Survival analysis can also be applied to the software itself, this has been demonstrated by Aman \emph{et al.} \cite{aman2017survival} and Caivano \emph{et al.} \cite{caivano2021exploratory}. 
Aman \emph{et al.} used survival analysis to analyze the time to a bug-fix for files modified by developers of different experience levels \cite{aman2017survival}. 
This analysis determined that files which were most recently modified by less experienced developers had an increased probability of needing a bug fix within a shorter time frame. 
Caivano \emph{et al.} explored the effect of dead code within OSS projects using survival analysis \cite{caivano2021exploratory}. 
They found that dead methods are present in Java code, persist for a long time before being buried or revived, are rarely revived, and that most dead methods have been dead since their inception. 
Other studies have examined the health of OSS projects using methods other than survival analysis. 
Xia \emph{et al.} predicted a number of health indicators of OSS projects, such as, the number of developers and the number of revisions.
These predictions were made using regression trees that were optimized using differential evolution, leading to a 10\% increase in prediction accuracy over the base line \cite{xia2020predicting}. 
Norick \emph{et al.} analyzed OSS projects using code quality measures and observed no significant evidence that the number of committing developers affects software quality \cite{norick2010effects}. 

\section{Data Set} \label{data}

Performing survival analysis of OSS projects requires a data set that records the repositories for projects on common hosting sites, including a history of all commits (referred to as revisions from here on) and major releases (revisions of note, often with a specific name and release date) \cite{ali2020cheating}. 
The \emph{popular-3k-python} subset~\footnote{https://annex.softwareheritage.org/public/dataset/graph/latest/popular-3k-python/sql/} of the Software Heritage Graph~\cite{pietri2019software} contains the necessary information and was used in the original paper and also in this paper.
This data set contains information on 3,052 popular Python projects hosted on GitHub/GitLab, Debian, and PyPI, and records revisions between 1980 and 2019 at the time of writing (the Software Heritage Graph is subject to updates, which makes it a non-reproducible data set).
Following the tutorial provided by the Software Heritage organization~\footnote{https://docs.softwareheritage.org/devel/swh-dataset/graph/postgresql.html}, a PostgreSQL database was hosted on the authors' local machines to facilitate data collection.

Though the \emph{popular-3k-python} data set contains all the necessary information to perform survival analysis, it first must be manipulated into a more suitable format before the analysis can be carried out. 
In this case, the collected data was manipulated such that the final data set contained the duration of the project, the censorship value, and the attributes of interest. 
Descriptions of each column present in the data set can be found in Table \ref{tab:data}. 
Data collection and manipulation were performed in Jupyter Notebooks, which are available in the replication package~\cite{repl-package}.

\begin{table*}
    \caption{Data Set Column Descriptions}
    \label{tab:data}
    \begin{tabular}{ll}
        \toprule
        Column Name & Description \\
        \midrule
        Duration\_months     & The duration of the project in months \\
        Censored             & True if the project's death is not observed (for more information see section \ref{death_censoring}) \\
        Host Type            & Which hosting service the project's repository resides on \\
        Major Releases       & True if the project publishes major releases \\
        High\_rev\_frequency & True if the project has a high revision frequency (greater than one revision per day) \\
        Multi\_repo          & True if the project is hosted on multiple hosting services\\
        High\_author\_count  & True if the project has a high author count (greater than twenty unique authors) \\
        \bottomrule
    \end{tabular}
\end{table*}

For their study, the original authors set a time frame of 165 months (where a month is defined as 28 days), starting in 2005 and ending in January 2018.
This paper uses the same time frame and determines exact start and end dates, as these were not given in the original paper.
Using January 1, 2018, as a strict end date and maintaining the study duration as 4,620 days (165 months as defined), the start date is found to be May 9, 2005.
After following the same procedures described in the original paper, a list of 2,066 projects and their associated information was obtained.

\section{Methods} \label{methods}

This section outlines how the data was transformed (Sections \ref{death_censoring} and \ref{revisionFreq}) and gives brief introductions to each of the statistical methods used (Sections \ref{surv_analysis} and \ref{bayes_surv_analysis}). 
All research artifacts can be found in the replication package \cite{repl-package}.

\subsection{Replication} \label{replmethods}
\subsubsection{Death and Censoring} \label{death_censoring}

Two critical concepts in survival analysis are events of interest and censoring.
As previously discussed, the event of interest for this study is project abandonment or death.
As defined in the introduction, a project is considered dead when it no longer receives any revisions.
With this definition, it is impossible to know whether any project is truly dead, because, unlike a living organism, any project may receive revisions at any point in the future, making it "not dead" by the working definition.
However, there are multiple practical ways of determining whether a project is dead, all of which rely on the scope of the studied data set.
For this study, the death of a project is determined by first defining two special revisions for each project: \emph{last revision} and \emph{last observed revision}. 
\emph{Last revision} is defined as the last recorded revision of a project within the scope of the data set (i.e., 1980 - 2019). 
\emph{Last observed revision} is defined as the last recorded revision of a project within the studied time frame (i.e., 2005 - 2018). 
If, and only if, the \emph{last revision} is also the \emph{last observed revision}, the project death is said to be observed. 

What happens if a project's death is not observed?
This is where censoring is used.
Suppose a project has its first revision one month before the end of the studied time frame, and continues to be revised daily past the studied time frame.
It would be incorrect to indicate that this project survived for only one month, but it is also impossible to include the project's future activity in the study.
Rather than discarding such projects, censoring causes them to be considered by the study for the observed duration, but without considering them as dead projects (i.e., they are removed from the calculations after they stop being observed).
There are multiple types of censoring in survival analysis, but this study uses random censoring or type III censoring, which involves removing subjects from a study at varying times relative to when they began to be observed (as is the case here) \cite{renganathan2016overview}. If a project's death is not observed, it is considered censored.

\begin{figure}[ht!]
    \centering
    \includegraphics[width=\columnwidth, keepaspectratio=true]{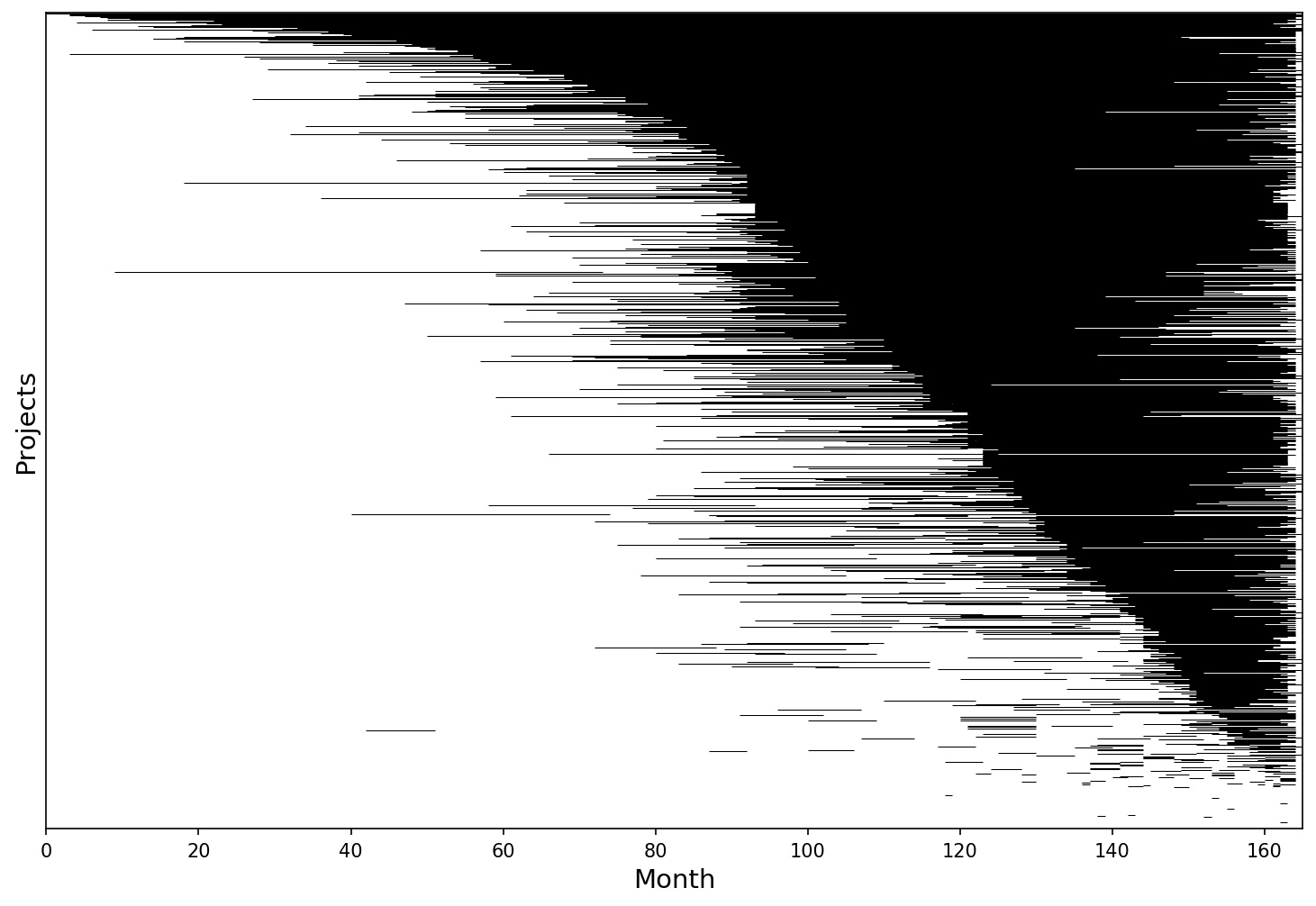}
    \caption{
        Project durations within the studied time frame.
        The projects are ordered by duration and plotted from and to their respective start and end dates.
        Month 0 begins on May 9, 2005, and Month 165 concludes on January 1, 2018.
        The black portion of the horizontal lines indicates the active period of a given project
        }
    \Description[Project durations within the studied time frame.]{Project durations within the studied time frame.
    The projects are ordered by duration and plotted from and to their respective start and end dates.
    Month 0 begins on May 9, 2005, and Month 165 concludes on January 1, 2018.
    The black portion of the horizontal lines indicates the active period of a given project}
    \label{fig:figure-1}
\end{figure}

The distribution of the project durations over the studied time frame can be seen in Figure \ref{fig:figure-1}, which is a replication of Figure 1 in the original paper.
Note that when the black lines extend to the end of the time frame, this likely indicates the project was censored.
Overall, about 62\% of the studied projects were censored.

\subsubsection{Survival Analysis} \label{surv_analysis}

Using both the calculated duration associated with each project and the censoring status of each project, the survival analysis can be performed.

The K-M estimator is a non-parametric estimation technique that estimates the survival function, $S(t)$.
The survival function gives the probability that a given project will survive until a particular time $t$.
At $t = 0$, the K-M estimator is 1 and as $t$ approaches infinity, the K-M estimator approaches 0.
More precisely, $S(t)$ is given by  $S(t) = p_0 \times p_1 \times p_2 \times \dots \times p_t$, where $p_i$ is the proportion of all projects that survived at the $i^{th}$ time step \cite{kaplan1958nonparametric}.
The K-M estimator produces curves that approach the true survival function of the data.
In other words, K-M curves plot the K-M survival probability against time.
The curve always goes in ``steps'' as the cumulative survival remains the same until the time of another event (in this case project abandonment). Furthermore, the censored observations are indicated by a vertical dash.

The hazard function is another useful function in survival analysis. It describes the probability of the event of interest or hazard (project abandonment in this case) occurring at a certain point in time given that the subject has survived to that time \cite{clark2003go}.
The original paper uses the Cox Proportional-Hazards model which allows for fitting a regression model in order to better understand how the health of projects relate to their key attributes. 
This analysis results in the hazards ratio (HR), which is derived from the model for all covariates that are included in the formula. 
Briefly, a $HR > 1$ indicates an increased risk of abandonment; on the other hand, $HR < 1$ indicates a decreased risk of abandonment \cite{cox1972regression}. 
As such, the HR represents a relative risk of abandonment that compares one instance of a binary feature (e.g. yes or no) to the other instance.

\subsection{Bayesian Survival Analysis} \label{bayes_surv_analysis}

The Bayesian approach to survival analysis is less common due to computational difficulties. 
However, it offers multiple advantages over the frequentist approach \cite{kelter2020bayesian}. 
This study replicates the methods outlined by Kelter \cite{kelter2020bayesian} to apply Bayesian survival analysis to the data set.
The Bayesian analysis uses posterior distributions of model parameters to draw inferences about them. 
These posterior distributions are obtained via Markov-Chain-Monte-Carlo (MCMC) algorithms. 
The statistical modelling language used was Stan.

A parametric exponential model that assumes the survival times $y = (y_1, y_2, \dots, y_n)$ for the set of projects are exponentially distributed with parameter $\lambda$ was created as shown in Equation \ref{eq:surv_times}.

\begin{equation} \label{eq:surv_times}
    f(y_i|\lambda) = \lambda\exp(- \lambda y_i) \mbox{ for } i=1,\dots,n
\end{equation}

The censoring indicators are denoted as $v = (v_1, v_2,\dots, v_n)$ where $v_i = 0$ if $y_i$ is right censored (project death is not observed) and $v_i = 1$ if $y_i$ is a failure time (project death is observed). The survival function, which is the probability of surviving past the time point $y_i$, is given by Equation \ref{eq:surv_func}.

\begin{equation} \label{eq:surv_func}
    S(y_i|\lambda) = P(T \geq yi|T \geq 0) = 1 - [1 - \exp(-  \lambda y_i)] = \exp(- \lambda y_i)
\end{equation}

The combination of Equations \ref{eq:surv_times} and \ref{eq:surv_func} yields the survival model given by Equations \ref{eq:model1} through \ref{eq:model4}.

\begin{equation} \label{eq:model1}
    y_i|v_i \sim f(y_i| \lambda)^{v_i} + S(y_i| \lambda)^{1-v_i} = [\lambda exp(-  \lambda y_i)]^{v_i} + [exp(-  \lambda y_i)]^{1-v_i}
\end{equation}
\begin{equation} \label{eq:model2}
    \lambda \sim p(\lambda)
\end{equation}
\begin{equation} \label{eq:model3}
    \lambda = exp(x_i^T \beta)
\end{equation}
\begin{equation} \label{eq:model4}
    \beta = normal(0, 10)
\end{equation}

This model was then used to visualize the posterior survival functions for the following five project attributes: major releases, hosting service of the project, use of multiple hosting services, team size, and revision frequency. 

\subsection{Revision Frequency Analysis} \label{revisionFreq}

The original paper mentions that ``The health of a project could be computed by the number and frequency of contributions...'' \cite{ali2020cheating} but does not directly study this measurement.
This paper seeks to explore the frequency of contributions as a method of assessment. 
Simply analyzing the number of contributions would not yield useful results given the varying nature of the project durations in the studied time frame. 
The revision frequency, defined as the number of commits divided by the number of days in the project's observed lifetime, was dichotomized into two groups depending on whether the frequency was above one revision per day.
Although the median revision frequency was approximately 0.68 revisions per day, the threshold value of one was chosen because it is easier to remember when keeping these attributes in mind and, similar to the other dichotomizing attributes, it provides a threshold that fewer projects attain to, which sets them apart.
This study applies both the frequentist and Bayesian analysis methods to the data to stratify the effects of high revision frequency on the overall health of an open-source project.

\section{Results} \label{results}

This section describes the results for each portion of the study. 
Section \ref{repl-results} describes the results for the replication portion of the study. 
Sections \ref{bayes-results} and \ref{rev-freq-results} describe the results of the Bayesian survival analysis and the revision frequency analysis respectively.

\subsection{Replication} \label{repl-results}

\begin{figure*}
    \centering
    \begin{subfigure}[b]{\columnwidth}
        \centering
        \includegraphics[width=\textwidth]{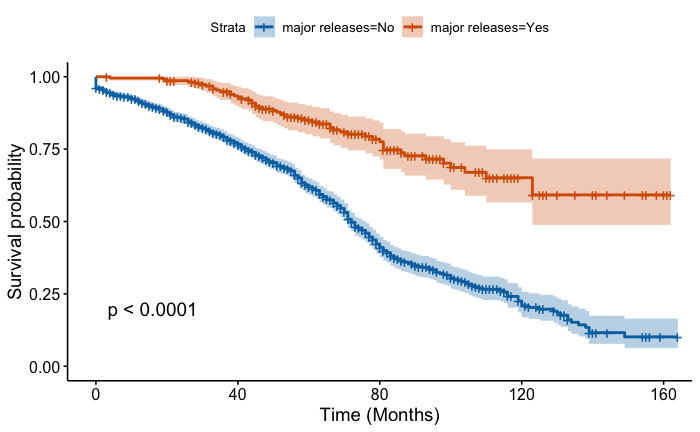}
        \caption{\small K-M curves for projects which publish major releases and those which do not} 
        \label{fig:major_releases}
    \end{subfigure}
    \hfill
    \begin{subfigure}[b]{\columnwidth}
        \centering 
        \includegraphics[width=\textwidth]{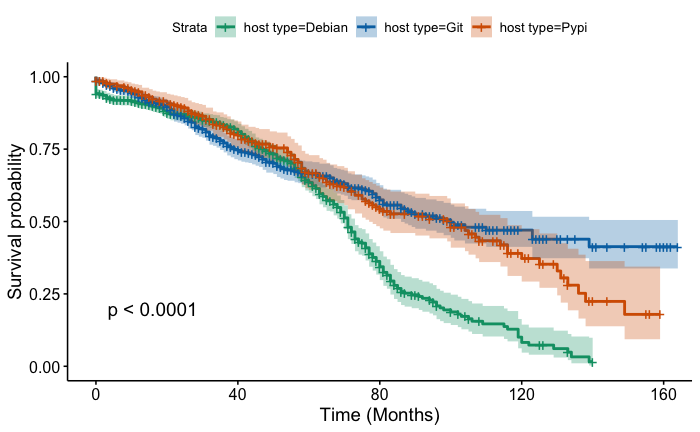}
        \caption{\small K-M curves for projects with different hosting services} 
        \label{fig:host_type}
    \end{subfigure}
    \vskip\baselineskip
    \begin{subfigure}[b]{\columnwidth}
        \centering 
        \includegraphics[width=\textwidth]{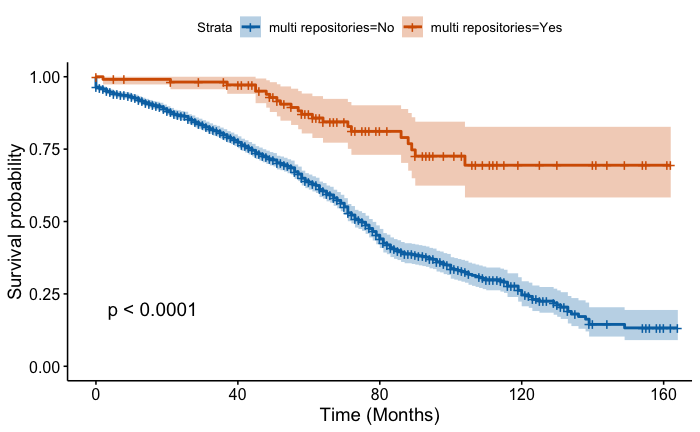}
        \caption{\small K-M curves for projects with repositories on multiple hosting services} 
        \label{fig:multi_repo}
    \end{subfigure}
    \hfill
    \begin{subfigure}[b]{\columnwidth} 
        \centering 
        \includegraphics[width=\textwidth]{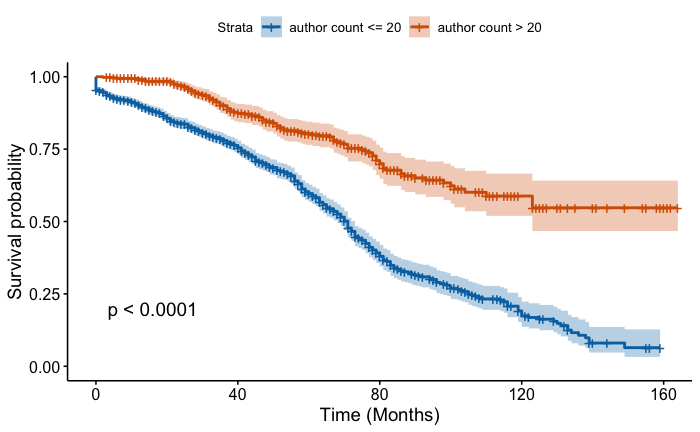}
        \caption{\small K-M curves for projects with high and low author count} 
        \label{fig:author_count}
    \end{subfigure}
    \caption{\small K-M curves for each project attribute} 
    \Description[Kaplan Meier curves for each project attribute]{Kaplan Meier curves for each project attribute}
    \label{fig:K-M curves}
\end{figure*}

\begin{figure*}
    \centering
    \includegraphics[scale=0.45, keepaspectratio=true]{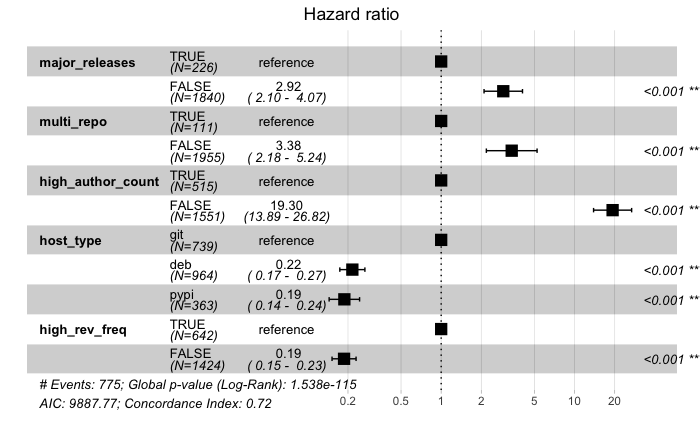}
    \caption{The Cox Proportional-Hazards model. From left to right: project attribute, attribute value with counts, hazard ratio, box plot of hazard ratio, and p-value of log rank test.}
    \Description[The Cox Proportional-Hazards model.]{The Cox Proportional-Hazards model. From left to right: project attribute, attribute value with counts, hazard ratio, box plot of hazard ratio, and p-value of log rank test.}
    \label{fig:Cox}
\end{figure*}

The replication study performed in this paper yielded extremely similar results to those presented in the original paper. 
Figure \ref{fig:K-M curves} depicts K-M curves along with their confidence intervals and p-values. 
The p-values imply that the difference in survival probability for projects within each group is statistically significant.
As seen in Figure \ref{fig:major_releases}, this study found that \emph{projects with at least one major release have higher chances of survival.} 
The curve for projects with at least one major release plateaus around 65\% survival probability around the 120-month mark, whereas the survival probability of projects with no major releases ends up less than 20\% by the end of the study period. 
Figure \ref{fig:host_type} represents the significance of the type of hosting service used. \emph{It was observed that projects that are hosted on GitHub have higher survival chances in the long run}, though the curves suggest all three hosting services have a similar trend for the first 55 months, which is within the average duration of projects hosted on these services.
In addition, \emph{having multiple repositories hosted on multiple services significantly increased the chances of a project's survival.}
As seen in Figure \ref{fig:multi_repo}, the survival rate for such projects is close to 70\% by the end of the study period, whereas it is around 20\% for projects with only one package repository system. 
The curve in Figure \ref{fig:author_count} illustrates the effect of the size of the network of developers, \emph{the projects with more than 20 different authors end up having a survival probability of about 60\% compared to a 20\% survival probability for projects with a small team of developers.
}
Figure \ref{fig:Cox} presents the results of the Cox Proportional-Hazards model which quantifies the effect of these attributes on project survival probability. 
The third column shows the hazard ratio which indicates the probability of abandonment with respect to the reference feature.
In the first row, the hazard ratio for projects which do not publish major releases with respect to projects that do, is nearly 3. 
This implies that \emph{the projects without major releases are three times more likely to become inactive compared to projects with at least one major release.} 
Similarly, \emph{projects with repositories on a single hosting service are 3.3 times more likely to be abandoned.}
The third row highlights that \emph{the projects with fewer developers are 19.3 times more likely to be inactive.}
For the type of hosting service used, the ratio implies that \emph{projects hosted on PyPI or Debian are less likely to be abandoned compared to projects that are hosted on GitHub.} 
This appears to contradict the results of the K-M curve and will be further discussed in Section \ref{discussion}.

\subsection{Bayesian Survival Analysis} \label{bayes-results}

\begin{figure*}
    \centering
    \begin{subfigure}[b]{\columnwidth}
        \centering
        \includegraphics[width=\textwidth]{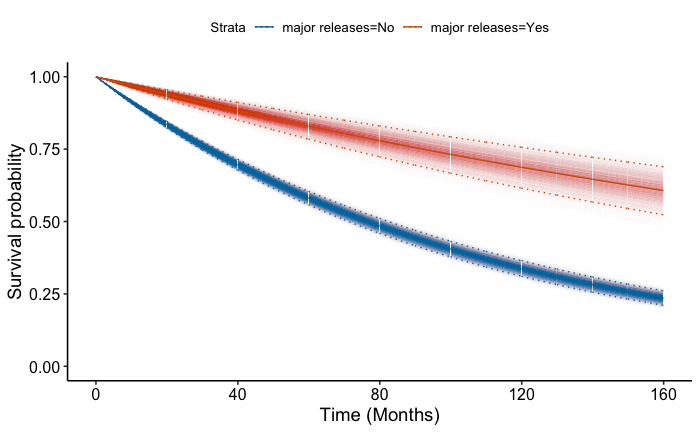}   
        \caption{\small Posterior survival functions for projects which publish major releases and those which do not}
        \label{fig:bayes_major_releases}
    \end{subfigure}
    \hfill
    \begin{subfigure}[b]{\columnwidth}
        \centering 
        \includegraphics[width=\textwidth]{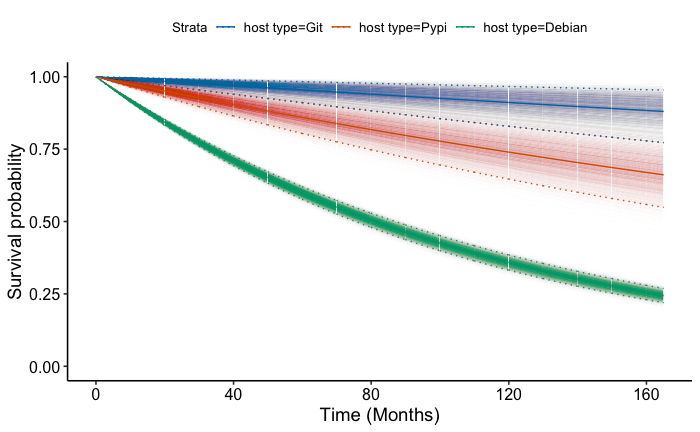}
        \caption{\small Posterior survival functions for projects with different hosting services}
        \label{fig:bayes_host_type}
    \end{subfigure}
    \vskip\baselineskip
    \begin{subfigure}[b]{\columnwidth}
        \centering 
        \includegraphics[width=\textwidth]{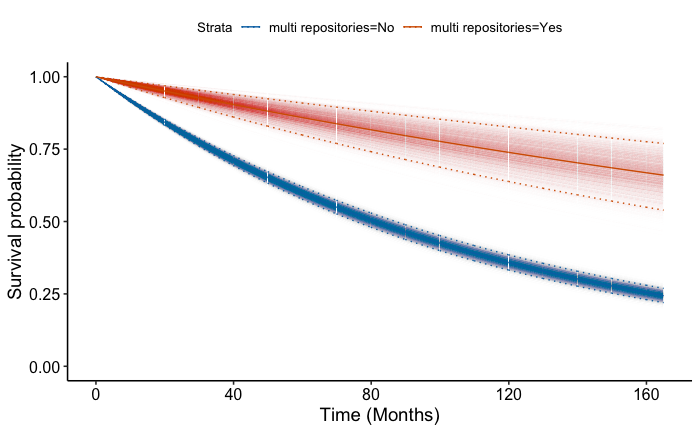}
        \caption{\small Posterior survival functions for projects with repositories on multiple hosting services}
        \label{fig:bayes_multi_repo}
    \end{subfigure}
    \hfill
    \begin{subfigure}[b]{\columnwidth}
        \centering 
        \includegraphics[width=\textwidth]{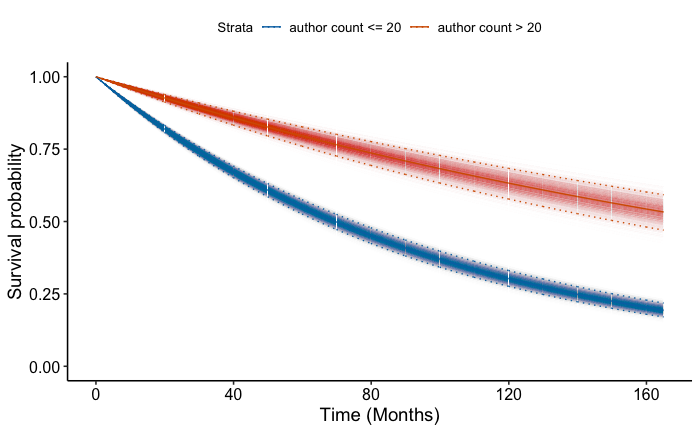} 
        \caption{\small Posterior survival functions for projects with high and low author count}
        \label{fig:bayes_author_count}
    \end{subfigure}
    \caption{\small Bayesian posterior survival functions for each project attribute} 
    \Description[Bayesian posterior survival functions for each project attribute]{Bayesian posterior survival functions for each project attribute}
    \label{fig:Bayesian posterior survival functions}
\end{figure*}

Figure \ref{fig:Bayesian posterior survival functions} shows the posterior survival functions for variations of the selected project attributes. 
The dotted lines represent the  2.5\% and 97.5\% quantiles, while the solid middle line represents the posterior mean of $\beta$ (the prior on the project attribute of interest). 
The remaining lines represent all valid posterior survival functions. 
Figure \ref{fig:bayes_major_releases} clearly indicates that the survival function of the projects with no major releases decreases much faster than projects with releases. 
\emph{Survival probability for projects with no major releases was lower than 25\% after 150 months compared to 55\% for projects with major releases.}
Figure \ref{fig:bayes_host_type} illustrates that \emph{projects hosted on Github have the highest survival chances compared to those hosted on PyPI and Debian.} 
At 150 months, the predicted survival probabilities for Github, PyPI and Debian were 87\%, 65\% and 25\%, respectively.
Figure \ref{fig:bayes_multi_repo} shows that \emph{projects with repositories on more than one hosting service had significantly higher survival chances than projects with repositories on a single hosting service.}
Above 65\% survival probability for multiple repositories compared to slightly over 25\% for single repository projects at the end of 150 months.
Figure \ref{fig:bayes_author_count} demonstrates the importance of the number of contributing developers on the survival chances of open-source python projects. 
For \emph{projects with more than 20 authors, the predicted survival probability was around 55\%, while less than 25\% for projects with less than 20 authors at the 150-month mark.}
Additionally, for all project attributes, the beta value for the 97.5\% quantile was smaller than zero, which ensures that the estimates for $\lambda$ have low uncertainty \cite{kelter2020bayesian}.

\subsection{Revision Frequency Analysis} \label{rev-freq-results}

\begin{figure*}
    \centering
    \begin{subfigure}[b]{\columnwidth}
        \centering
        \includegraphics[width=\columnwidth, keepaspectratio=true]{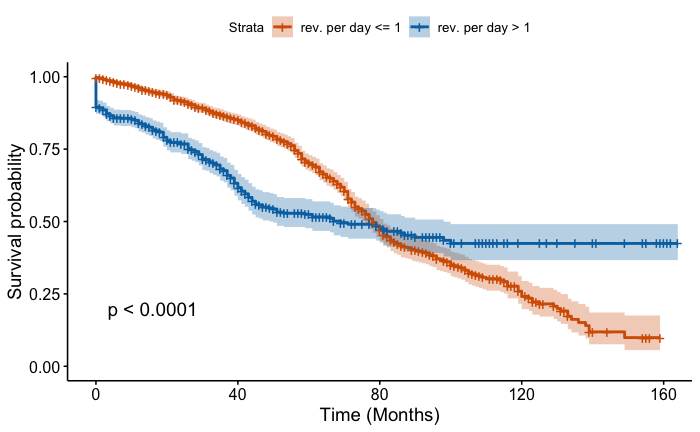}  
        \caption{}
        \label{fig:K-M_curve_for_revision_frequency}
    \end{subfigure}
    \hfill
    \begin{subfigure}[b]{\columnwidth}
        \centering 
        \includegraphics[width=\columnwidth, keepaspectratio=true]{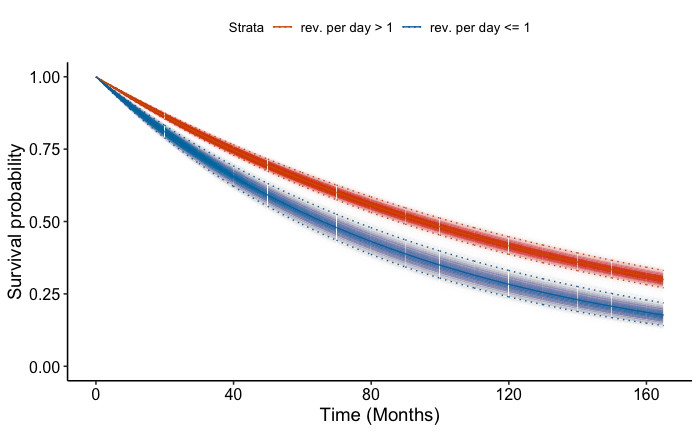}
        \caption{}
        \label{fig:Bayesian_curve_for_revision_frequency}
    \end{subfigure}
    \caption{\small K-M (a) and Bayesian posterior survival function (b) curves for revision frequency} 
    \Description[Kaplan Meier and Bayesian posterior survival function curves for revision frequency]{Kaplan Meier and Bayesian posterior survival function curves for revision frequency}
    \label{fig:Revision frequency}
\end{figure*}

The K-M curves for revision frequency are different from the graphs generated for the other attributes, as seen in Figure \ref{fig:K-M_curve_for_revision_frequency}. 
A curve drop was observed for the projects with more than one revision per day in the beginning of the study period.
In addition, the projects with lower revision frequencies out performed projects with higher revision frequencies during the mean duration period of the studied projects.
Figure \ref{fig:Bayesian_curve_for_revision_frequency} shows the posterior survival functions for projects with more than one revision per day and those with revision per day less than or equal to one. Although a significant difference was not observed in predicted survival probabilities, projects with more than one revision per day had a slightly higher chance of survival by the end of the study period.
As shown in Figure \ref{fig:Cox}, the results of the regression model indicate that the projects with a higher revision frequency are more likely to be abandoned.

\section{Discussion} \label{discussion}

The replication was deemed successful in that the results were consistent with the qualitative findings of the original authors with respect to the first research question.
As in other works \cite{samoladas2010survival}, the number of developers was found to be a significant indicator of a project's health in terms of duration (note the smaller confidence interval for Figure \ref{fig:author_count}), and this result lends support to "Linus's Law", which claims that a larger number of developers will make bug detection and resolution simpler \cite{raymond2001LinusLaw}, though this study does not make claims about a project's quality.
Interestingly, Norick \emph{et al.} observed no significant evidence that software quality is affected by the number of developers \cite{norick2010effects}.
This suggests a higher number of developers has an effect on the longevity of a project that is unrelated to the quality of the software.

The Cox Proportional-Hazards model operates under the assumption that the hazard ratio between the hazard functions of each group remains the same throughout the period of the study \cite{stel2011cox}.
Looking at the K-M curves for this study, it can be concluded that the proportional hazards assumption does not hold as the survival functions diverge over time and cross over each other rather than running in parallel \cite{persson2007ACO}.
This means the hazard ratios are not applicable to the entire studied time frame.
After inspecting the hazard ratios in Figure \ref{fig:Cox}, it is assumed the calculations in the Cox model give more weight to the portion of the study containing a larger number of subjects (i.e., the left most portion of the K-M curves).
This partially explains why there is such a large discrepancy between the hazard ratio for the author count attribute found in this paper and the one found in the original paper (19.3 compared to 5.95).

When comparing the results of the Bayesian analysis to those of the frequentist analysis, it is clear that the Bayesian survival functions reveal less information about the data set.
This is to be expected, however, as the models require the data to conform to the chosen prior distribution function and cannot reflect more complex detail.
This comparison is especially apparent when looking at Figure \ref{fig:Revision frequency}.
One of the advantages of Bayesian models is their ability to create predictions in the presence of new data.
Whereas the frequentist approach attempts to reveal exactly what the data set is describing, this would be considered overfitting in the context of Bayesian analysis.
Though the approaches are distinct and serve different purposes, it is clear that the survival curves generated by both approaches share the same sentiment with respect to the project attributes.

The survival functions displayed in Figure \ref{fig:Revision frequency} are distinct when compared with the other results of the study.
There is a noticeable drop in the K-M curve belonging to projects with high revision frequencies within the first few months of the studied time period.
This is likely due to the presence of many short-lived projects with rapid development times.
If such projects were removed from the study, the projects with high revision frequencies would perform even better.
Additionally, it can be seen that, within the mean project duration (about 46 months), the projects with higher revision frequencies are abandoned more frequently, but after this point their curve begins to plateau while the other curve begins to decline more rapidly.
This suggests that \emph{having a higher revision frequency benefits a project's probability of survival in the long run and may indicate that consistency is more important}, though the metric used for this study does not consider how evenly distributed the revisions of a project are.

Finally, it is hypothesized from the K-M curves of every binary project attribute (Figures \ref{fig:K-M curves} and \ref{fig:Revision frequency}) that somewhere around the 60 month point represents a critical threshold for projects, after which, the attributes outlined in this study have an increasingly significant effect.
More work should be done to investigate whether this is truly the case.

\subsection{Implications} \label{implications}

\subsubsection{Implications for Software Practitioners} 

The results of this study, and the studies it is based on, have practical implications.
They give developers of open-source software insights into which projects are likely to receive long term attention from other developers, they help project coordinators understand the attributes that their projects should possess if they want a long-lasting project, and they give companies and organizations looking to invest in, utilize, or rely on OSS projects confidence that such projects will remain active.

\subsubsection{Implications for Researchers} 

Though this paper's results were qualitatively similar to those of the original authors, it is suspected that a variation in approaches for data retrieval and the analysis tools used caused quantitative discrepancies (as seen in the hazard ratio for the author count attribute).
The authors of this paper therefore call for researchers to improve the reproducibility of their studies by providing research artifacts.

The application of Bayesian survival analysis presented in this paper is meant to serve as a preliminary investigation of the approach in the context of software duration.
It is suggested that researchers looking to use this method in the future take a deeper look into priors, likelihoods, and covariates.
Researchers may also wish to test the validity of the models produced in this study by using them to predict software duration for new data sets.

\subsection{Limitations} \label{limit}

\subsubsection{Limitations of the Methods} \label{limit-methods}

The original paper \cite{ali2020cheating} and the corresponding presentation given \cite{ali2020video} have seemingly contradicting methods of censoring.
The method discussed in the original paper was deemed superior and is used in this paper.


Being a replication, this study was limited to choosing the methods used by the original paper.
These methods are also unanimous in the area of survival analysis, and there are few alternatives to them.
This study attempts to address these limitations by providing an alternative analysis method, namely Bayesian analysis.

When applying the K-M estimator, it is common to use a log-rank test to test the significance between the two groups which are being compared.
The log-rank test only indicates whether or not the probability of survival is statistically significant between the two groups and is not able to provide any information about the size of the difference between the two groups \cite{stel2011kaplan}.
Additionally, the K-M estimator does not account for confounding factors \cite{stel2011kaplan}.
In more traditional uses of the K-M estimator, an example of a confounding factor could be the age of the study participants.
In the case of this study, there may be confounding factors such as the experience level of the developers or whether the developers received funding to work on the project.
Neither of these factors are represented in the data set.

As discussed in the start of Section \ref{discussion}, the Cox Proportional-Hazards model is used with the assumption that, over the period of observation, the hazards within each group are proportional.
If the assumption is not true, then the Cox Proportional-Hazards model will lead to incorrect estimates of the hazard ratio between two groups \cite{stel2011cox}.
The assumption does not hold in this study, which explains the discrepancies between the results of the Cox regression model and the K-M curves.
It is unclear whether these assumptions are reasonable in the context of software duration.
Future studies should perform tests to determine whether the assumptions for their models hold and should seek methods for mitigating such errors through identifying time-dependant covariates.

The Bayesian approach to survival analysis comes with its own limitations as well.
As pointed out by Renganathan, Bayesian survival analysis can be subjective as the analyst places their own bias into the model when selecting the prior distributions \cite{renganathan2016overview}.
To mitigate this bias, prior selection requires both epistemological and ontological reasoning.
Prior distributions were chosen based on survival analysis done in other domains \cite{kelter2020bayesian, rethinking}, but more investigation should be done as to whether these priors are appropriate for this domain and whether the models from this study accurately predict durations of different data sets.

\subsubsection{Limitations of the Data}

The data set in this study has been aggregated from multiple version control systems across the web over a large period of time.
As such, the data set is not fully reproducible, as pointed out by the authors of the Software Heritage organization \cite{pietri2019software}.
Additionally, it cannot be ensured that the data contains a full history of the respective repositories.
The lack of certainty about the full history is because the repository admin can modify the history of revisions to suit their liking \cite{perils2009}.

There are inherent differences in the ways developers use the different hosting services.
Traditionally, services such as PyPI and Debian are used to host major releases of a product.
This may hide information about the number of developers and the revision frequency.
Additionally, the potential confounding factors mentioned in section \ref{limit-methods} are not represented in this data set.

It may also be worth noting that this data is only for Python projects and it is possible that different behaviours are associated with development in different languages.
Python is a relatively easy language to use, and can often be used for small tasks that are not maintained.
The results of the revision frequency analysis in section \ref{rev-freq-results} seem to indicate a large number of short lived projects.

The data set contains a large portion of censored data.
This means the abandonment of most of the projects was not observed.
As data points are censored (denoted by the vertical tick marks in the K-M curves), there is a smaller and smaller group of data points to study.
This means that the results towards the 165 month mark may be less representative. Finally, the data set contained many revisions (over 4 million) that were not associated with project URLs, the cause of this is unclear.

\subsection{Future Work} \label{future}

It can be said with high confidence that this is the first application of Bayesian survival analysis in the context of OSS. 
As such, prior distributions were uninformative and priors and likelihoods alike were chosen based on the application of Bayesian survival analysis done in other domains. 
More investigation should be conducted to determine if the priors and likelihoods applied in this paper are suitable for OSS. 
Furthermore, future work should investigate the predictive power of the Bayesian survival models presented here to determine their performance on other data sets. 
Similarly, an evaluation of the assumptions present in the Cox Proportional-Hazards model should be performed to determine if these assumptions are reasonable for the domain of OSS.

In regards to the study design, future work should increase the time frame of the study to determine how the attributes analyzed in this paper play out over a longer time period. 
Similarly, as only OSS written in Python was studied here, future work should evaluate the survival of OSS written in other popular languages such as Java. 

As noted in the discussion, a higher number of developers has some effect on the longevity of a project that is not related to the quality of the software being developed. 
Future works should investigate the reason behind this increased longevity and attempt to determine a reason for it. 
In line with Samoladas \emph{et al.} \cite{samoladas2010survival}, distinguishing between core developers and sporadic ones could lead to an increased understanding of the aforementioned phenomena. 

\section{Conclusion} \label{conclusion}

The work presented in this paper aimed to replicate that of the original authors \cite{ali2020cheating}. 
In addition, one more project attribute was analyzed, the effect of revision frequency on open-source software project duration.
Furthermore, Bayesian survival analysis was used to study the attributes in the original paper as well as revision frequency.
While the Kaplan Meier curves are consistent with those of the original paper, the Cox Proportional-Hazards Model indicated a quantitative discrepancy with regard to the author count hazard ratio.
This discrepancy is likely due to a difference in methods which could have been mitigated through the use of research artifacts.
It was observed that a higher revision frequency leads to a better chance of survival by the end of the study duration. 
The Bayesian survival analysis yielded similar results to the frequentist approach.
However, it is worth noting that the results obtained from the Bayesian approach were much less granular and are intended to predict new data rather than accurately describe the studied data.

\begin{acks}
    We sincerely thank Dr. Neil Ernst and the members of the CHISEL Group for their support and guidance throughout this research project. 
\end{acks}
\balance
\bibliographystyle{ACM-Reference-Format}
\bibliography{report}


\begin{thebibliography}{00}


\ifx \showCODEN    \undefined \def \showCODEN     #1{\unskip}     \fi
\ifx \showDOI      \undefined \def \showDOI       #1{#1}\fi
\ifx \showISBNx    \undefined \def \showISBNx     #1{\unskip}     \fi
\ifx \showISBNxiii \undefined \def \showISBNxiii  #1{\unskip}     \fi
\ifx \showISSN     \undefined \def \showISSN      #1{\unskip}     \fi
\ifx \showLCCN     \undefined \def \showLCCN      #1{\unskip}     \fi
\ifx \shownote     \undefined \def \shownote      #1{#1}          \fi
\ifx \showarticletitle \undefined \def \showarticletitle #1{#1}   \fi
\ifx \showURL      \undefined \def \showURL       {\relax}        \fi
\providecommand\bibfield[2]{#2}
\providecommand\bibinfo[2]{#2}
\providecommand\natexlab[1]{#1}
\providecommand\showeprint[2][]{arXiv:#2}

\bibitem[\protect\citeauthoryear{??}{rep}{2022}]%
        {repl-package}
 \bibinfo{year}{2022}\natexlab{}.
\newblock \showarticletitle{{Two Approaches to Survival Analysis of Open Source
  Python Projects}}. \bibinfo{publisher}{Zenodo}.
\newblock
\showDOI{%
\url{https://doi.org/10.5281/zenodo.6040657}}


\bibitem[\protect\citeauthoryear{Ali, Parlett-Pelleriti, and Linstead}{Ali
  et~al\mbox{.}}{2020a}]%
        {ali2020cheating}
\bibfield{author}{\bibinfo{person}{Rao~Hamza Ali}, \bibinfo{person}{Chelsea
  Parlett-Pelleriti}, {and} \bibinfo{person}{Erik Linstead}.}
  \bibinfo{year}{2020}\natexlab{a}.
\newblock \showarticletitle{Cheating Death: A Statistical Survival Analysis of
  Publicly Available Python Projects}. In \bibinfo{booktitle}{{\em Proceedings
  of the 17th International Conference on Mining Software Repositories}}.
  \bibinfo{publisher}{Association for Computing Machinery},
  \bibinfo{address}{New York, NY, USA}, \bibinfo{pages}{6--10}.
\newblock
\showISBNx{9781450375177}
\showURL{%
\url{https://doi.org/10.1145/3379597.3387511}}


\bibitem[\protect\citeauthoryear{Ali, Parlett-Pelleriti, and Linstead}{Ali
  et~al\mbox{.}}{2020b}]%
        {ali2020video}
\bibfield{author}{\bibinfo{person}{Rao~Hamza Ali}, \bibinfo{person}{Chelsea
  Parlett-Pelleriti}, {and} \bibinfo{person}{Erik Linstead}.}
  \bibinfo{year}{2020}\natexlab{b}.
\newblock \bibinfo{title}{Cheating Death: A Statistical Survival Analysis of
  Publicly Available Python Projects (MSR 2020 - Mining Challenge) - MSR 2020}.
\newblock
  \bibinfo{howpublished}{https://2020.msrconf.org/details/msr-2020-mining-challenge/1/Cheating-Death-A-Statistical-Survival-Analysis-of-Publicly-Available-Python-Projects}.
    (\bibinfo{date}{June} \bibinfo{year}{2020}).
\newblock
\newblock
\shownote{(Accessed on 11/17/2021).}


\bibitem[\protect\citeauthoryear{Aman, Amasaki, Yokogawa, and Kawahara}{Aman
  et~al\mbox{.}}{2017}]%
        {aman2017survival}
\bibfield{author}{\bibinfo{person}{Hirohisa Aman}, \bibinfo{person}{Sousuke
  Amasaki}, \bibinfo{person}{Tomoyuki Yokogawa}, {and} \bibinfo{person}{Minoru
  Kawahara}.} \bibinfo{year}{2017}\natexlab{}.
\newblock \showarticletitle{A Survival Analysis of Source Files Modified by New
  Developers}. In \bibinfo{booktitle}{{\em Product-Focused Software Process
  Improvement}}, \bibfield{editor}{\bibinfo{person}{Michael Felderer},
  \bibinfo{person}{Daniel M{\'e}ndez~Fern{\'a}ndez}, \bibinfo{person}{Burak
  Turhan}, \bibinfo{person}{Marcos Kalinowski}, \bibinfo{person}{Federica
  Sarro}, {and} \bibinfo{person}{Dietmar Winkler}} (Eds.).
  \bibinfo{publisher}{Springer International Publishing},
  \bibinfo{address}{Innsbruck, Austria}, \bibinfo{pages}{80--88}.
\newblock
\showISBNx{978-3-319-69926-4}


\bibitem[\protect\citeauthoryear{Bird, Rigby, Barr, Hamilton, German, and
  Devanbu}{Bird et~al\mbox{.}}{2009}]%
        {perils2009}
\bibfield{author}{\bibinfo{person}{Christian Bird}, \bibinfo{person}{Peter~C
  Rigby}, \bibinfo{person}{Earl~T Barr}, \bibinfo{person}{David~J Hamilton},
  \bibinfo{person}{Daniel~M German}, {and} \bibinfo{person}{Prem Devanbu}.}
  \bibinfo{year}{2009}\natexlab{}.
\newblock \showarticletitle{The promises and perils of mining git}. In
  \bibinfo{booktitle}{{\em 2009 6th IEEE International Working Conference on
  Mining Software Repositories}}. \bibinfo{publisher}{IEEE},
  \bibinfo{address}{Vancouver, Canada}, \bibinfo{pages}{1--10}.
\newblock


\bibitem[\protect\citeauthoryear{Caivano, Cassieri, Romano, and
  Scanniello}{Caivano et~al\mbox{.}}{2021}]%
        {caivano2021exploratory}
\bibfield{author}{\bibinfo{person}{Danilo Caivano}, \bibinfo{person}{Pietro
  Cassieri}, \bibinfo{person}{Simone Romano}, {and} \bibinfo{person}{Giuseppe
  Scanniello}.} \bibinfo{year}{2021}\natexlab{}.
\newblock \showarticletitle{An Exploratory Study on Dead Methods in Open-Source
  Java Desktop Applications}. In \bibinfo{booktitle}{{\em Proceedings of the
  15th ACM / IEEE International Symposium on Empirical Software Engineering and
  Measurement (ESEM)}}. \bibinfo{publisher}{Association for Computing
  Machinery}, \bibinfo{address}{New York, NY, USA}, Article
  \bibinfo{articleno}{10}, \bibinfo{numpages}{11}~pages.
\newblock
\showISBNx{9781450386654}
\showURL{%
\url{https://doi.org/10.1145/3475716.3475773}}


\bibitem[\protect\citeauthoryear{Clark, Bradburn, Love, and Altman}{Clark
  et~al\mbox{.}}{2003}]%
        {clark2003go}
\bibfield{author}{\bibinfo{person}{Taane~G Clark}, \bibinfo{person}{Michael~J
  Bradburn}, \bibinfo{person}{Sharon~B Love}, {and} \bibinfo{person}{Douglas~G
  Altman}.} \bibinfo{year}{2003}\natexlab{}.
\newblock \showarticletitle{Survival analysis part I: basic concepts and first
  analyses}.
\newblock \bibinfo{journal}{{\em British journal of cancer\/}}
  \bibinfo{volume}{89}, \bibinfo{number}{2} (\bibinfo{year}{2003}),
  \bibinfo{pages}{232--238}.
\newblock


\bibitem[\protect\citeauthoryear{Cox}{Cox}{1972}]%
        {cox1972regression}
\bibfield{author}{\bibinfo{person}{David~R Cox}.}
  \bibinfo{year}{1972}\natexlab{}.
\newblock \showarticletitle{Regression models and life-tables}.
\newblock \bibinfo{journal}{{\em Journal of the Royal Statistical Society:
  Series B (Methodological)\/}} \bibinfo{volume}{34}, \bibinfo{number}{2}
  (\bibinfo{year}{1972}), \bibinfo{pages}{187--202}.
\newblock


\bibitem[\protect\citeauthoryear{Evangelopoulos, Sidorova, Fotopoulos, and
  Chengalur-Smith}{Evangelopoulos et~al\mbox{.}}{2008}]%
        {evangelopoulos}
\bibfield{author}{\bibinfo{person}{Nicholas Evangelopoulos},
  \bibinfo{person}{Anna Sidorova}, \bibinfo{person}{Stergios Fotopoulos}, {and}
  \bibinfo{person}{Indushobha Chengalur-Smith}.}
  \bibinfo{year}{2008}\natexlab{}.
\newblock \showarticletitle{Determining Process Death Based on Censored
  Activity Data}.
\newblock \bibinfo{journal}{{\em Communications in Statistics - Simulation and
  Computation\/}} \bibinfo{volume}{37}, \bibinfo{number}{8}
  (\bibinfo{year}{2008}), \bibinfo{pages}{1647--1662}.
\newblock
\showDOI{%
\url{https://doi.org/10.1080/03610910802140224}}
\showeprint{https://doi.org/10.1080/03610910802140224}


\bibitem[\protect\citeauthoryear{Kaplan and Meier}{Kaplan and Meier}{1958}]%
        {kaplan1958nonparametric}
\bibfield{author}{\bibinfo{person}{E.~L. Kaplan} {and} \bibinfo{person}{Paul
  Meier}.} \bibinfo{year}{1958}\natexlab{}.
\newblock \showarticletitle{Nonparametric Estimation from Incomplete
  Observations}.
\newblock \bibinfo{journal}{{\it J. Amer. Statist. Assoc.}}
  \bibinfo{volume}{53}, \bibinfo{number}{282} (\bibinfo{year}{1958}),
  \bibinfo{pages}{457--481}.
\newblock
\showDOI{%
\url{https://doi.org/10.1080/01621459.1958.10501452}}
\showeprint{https://www.tandfonline.com/doi/pdf/10.1080/01621459.1958.10501452}


\bibitem[\protect\citeauthoryear{Kelter}{Kelter}{2020}]%
        {kelter2020bayesian}
\bibfield{author}{\bibinfo{person}{Riko Kelter}.}
  \bibinfo{year}{2020}\natexlab{}.
\newblock \showarticletitle{Bayesian survival analysis in STAN for improved
  measuring of uncertainty in parameter estimates}.
\newblock \bibinfo{journal}{{\em Measurement: Interdisciplinary Research and
  Perspectives\/}} \bibinfo{volume}{18}, \bibinfo{number}{2}
  (\bibinfo{year}{2020}), \bibinfo{pages}{101--109}.
\newblock


\bibitem[\protect\citeauthoryear{Lin, Robles, and Serebrenik}{Lin
  et~al\mbox{.}}{2017}]%
        {lin2017developer}
\bibfield{author}{\bibinfo{person}{Bin Lin}, \bibinfo{person}{Gregorio Robles},
  {and} \bibinfo{person}{Alexander Serebrenik}.}
  \bibinfo{year}{2017}\natexlab{}.
\newblock \showarticletitle{Developer Turnover in Global, Industrial Open
  Source Projects: Insights from Applying Survival Analysis}. In
  \bibinfo{booktitle}{{\em 2017 IEEE 12th International Conference on Global
  Software Engineering (ICGSE)}}. \bibinfo{publisher}{IEEE},
  \bibinfo{address}{Buenos Aires, Argentina}, \bibinfo{pages}{66--75}.
\newblock
\showDOI{%
\url{https://doi.org/10.1109/ICGSE.2017.11}}


\bibitem[\protect\citeauthoryear{McElreath}{McElreath}{2015}]%
        {rethinking}
\bibfield{author}{\bibinfo{person}{Richard McElreath}.}
  \bibinfo{year}{2020;2016;2015;}\natexlab{}.
\newblock \bibinfo{booktitle}{{\em Statistical rethinking: a Bayesian course
  with examples in R and Stan\/} (\bibinfo{edition}{second;1;} ed.)}.
  Vol.~\bibinfo{volume}{122}.
\newblock \bibinfo{publisher}{Chapman \& Hall/CRC}, \bibinfo{address}{Boca
  Raton}.
\newblock
\showISBNx{0429642318;9780429635977;9780429029608;036713991X;9780367139919;0429639147;9780429642319;0429635974;9780429639142;0429029608;1482253445;9781482253443;}


\bibitem[\protect\citeauthoryear{Miller, Widder, K{\"a}stner, and
  Vasilescu}{Miller et~al\mbox{.}}{2019}]%
        {miller2019people}
\bibfield{author}{\bibinfo{person}{Courtney Miller},
  \bibinfo{person}{David~Gray Widder}, \bibinfo{person}{Christian K{\"a}stner},
  {and} \bibinfo{person}{Bogdan Vasilescu}.} \bibinfo{year}{2019}\natexlab{}.
\newblock \showarticletitle{Why Do People Give Up FLOSSing? A Study of
  Contributor Disengagement in Open Source}. In \bibinfo{booktitle}{{\em Open
  Source Systems}}, \bibfield{editor}{\bibinfo{person}{Francis Bordeleau},
  \bibinfo{person}{Alberto Sillitti}, \bibinfo{person}{Paulo Meirelles}, {and}
  \bibinfo{person}{Valentina Lenarduzzi}} (Eds.). \bibinfo{publisher}{Springer
  International Publishing}, \bibinfo{address}{Cham},
  \bibinfo{pages}{116--129}.
\newblock
\showISBNx{978-3-030-20883-7}


\bibitem[\protect\citeauthoryear{Norick, Krohn, Howard, Welna, and
  Izurieta}{Norick et~al\mbox{.}}{2010}]%
        {norick2010effects}
\bibfield{author}{\bibinfo{person}{Brandon Norick}, \bibinfo{person}{Justin
  Krohn}, \bibinfo{person}{Eben Howard}, \bibinfo{person}{Ben Welna}, {and}
  \bibinfo{person}{Clemente Izurieta}.} \bibinfo{year}{2010}\natexlab{}.
\newblock \showarticletitle{Effects of the Number of Developers on Code Quality
  in Open Source Software: A Case Study}. In \bibinfo{booktitle}{{\em
  Proceedings of the 2010 ACM-IEEE International Symposium on Empirical
  Software Engineering and Measurement}} {\em (\bibinfo{series}{ESEM '10})}.
  \bibinfo{publisher}{Association for Computing Machinery},
  \bibinfo{address}{New York, NY, USA}, Article \bibinfo{articleno}{62},
  \bibinfo{numpages}{1}~pages.
\newblock
\showISBNx{9781450300391}
\showDOI{%
\url{https://doi.org/10.1145/1852786.1852864}}


\bibitem[\protect\citeauthoryear{Ortega and Izquierdo-Cortazar}{Ortega and
  Izquierdo-Cortazar}{2009}]%
        {ortega2009survival}
\bibfield{author}{\bibinfo{person}{Felipe Ortega} {and} \bibinfo{person}{Daniel
  Izquierdo-Cortazar}.} \bibinfo{year}{2009}\natexlab{}.
\newblock \showarticletitle{Survival analysis in open development projects}. In
  \bibinfo{booktitle}{{\em 2009 ICSE Workshop on Emerging Trends in
  Free/Libre/Open Source Software Research and Development}}.
  \bibinfo{publisher}{IEEE}, \bibinfo{address}{Washington, DC, USA},
  \bibinfo{pages}{7--12}.
\newblock
\showDOI{%
\url{https://doi.org/10.1109/FLOSS.2009.5071353}}


\bibitem[\protect\citeauthoryear{Persson and Khamis}{Persson and
  Khamis}{2007}]%
        {persson2007ACO}
\bibfield{author}{\bibinfo{person}{Inger Persson} {and}
  \bibinfo{person}{Harry~J Khamis}.} \bibinfo{year}{2007}\natexlab{}.
\newblock \showarticletitle{A comparison of graphical methods for assessing the
  proportional hazards assumptions in the Cox model}.
\newblock \bibinfo{journal}{{\em Journal of Statistics and Applications\/}}
  \bibinfo{volume}{2}, \bibinfo{number}{1-4} (\bibinfo{year}{2007}),
  \bibinfo{pages}{1}.
\newblock


\bibitem[\protect\citeauthoryear{Pietri, Spinellis, and Zacchiroli}{Pietri
  et~al\mbox{.}}{2019}]%
        {pietri2019software}
\bibfield{author}{\bibinfo{person}{Antoine Pietri}, \bibinfo{person}{Diomidis
  Spinellis}, {and} \bibinfo{person}{Stefano Zacchiroli}.}
  \bibinfo{year}{2019}\natexlab{}.
\newblock \showarticletitle{The Software Heritage Graph Dataset: Public
  Software Development Under One Roof}. In \bibinfo{booktitle}{{\em 2019
  IEEE/ACM 16th International Conference on Mining Software Repositories
  (MSR)}}. \bibinfo{publisher}{IEEE}, \bibinfo{address}{Montreal, QC, Canada},
  \bibinfo{pages}{138--142}.
\newblock
\showDOI{%
\url{https://doi.org/10.1109/MSR.2019.00030}}


\bibitem[\protect\citeauthoryear{Raymond}{Raymond}{2001}]%
        {raymond2001LinusLaw}
\bibfield{author}{\bibinfo{person}{Eric~S. Raymond}.}
  \bibinfo{year}{2001}\natexlab{}.
\newblock \bibinfo{booktitle}{{\em The cathedral and the bazaar: musings on
  Linux and Open Source by an accidental revolutionary\/}
  (\bibinfo{edition}{rev.} ed.)}.
\newblock \bibinfo{publisher}{O'Reilly}, \bibinfo{address}{Beijing;Cambridge,
  Mass;}.
\newblock
\showISBNx{0596001312;0596001088;9780596001315;9780596001087;}


\bibitem[\protect\citeauthoryear{Renganathan}{Renganathan}{2016}]%
        {renganathan2016overview}
\bibfield{author}{\bibinfo{person}{Vinaitheerthan Renganathan}.}
  \bibinfo{year}{2016}\natexlab{}.
\newblock \showarticletitle{Overview of frequentist and bayesian approach to
  survival analysis}.
\newblock \bibinfo{journal}{{\em Applied Medical Informatics.\/}}
  \bibinfo{volume}{38}, \bibinfo{number}{1} (\bibinfo{year}{2016}),
  \bibinfo{pages}{25--38}.
\newblock


\bibitem[\protect\citeauthoryear{Samoladas, Angelis, and Stamelos}{Samoladas
  et~al\mbox{.}}{2010}]%
        {samoladas2010survival}
\bibfield{author}{\bibinfo{person}{Ioannis Samoladas},
  \bibinfo{person}{Lefteris Angelis}, {and} \bibinfo{person}{Ioannis
  Stamelos}.} \bibinfo{year}{2010}\natexlab{}.
\newblock \showarticletitle{Survival analysis on the duration of open source
  projects}.
\newblock \bibinfo{journal}{{\em Information and Software Technology\/}}
  \bibinfo{volume}{52}, \bibinfo{number}{9} (\bibinfo{year}{2010}),
  \bibinfo{pages}{902--922}.
\newblock


\bibitem[\protect\citeauthoryear{Shull, Carver, Vegas, and Juristo}{Shull
  et~al\mbox{.}}{2008}]%
        {shull2008role}
\bibfield{author}{\bibinfo{person}{Forrest~J Shull}, \bibinfo{person}{Jeffrey~C
  Carver}, \bibinfo{person}{Sira Vegas}, {and} \bibinfo{person}{Natalia
  Juristo}.} \bibinfo{year}{2008}\natexlab{}.
\newblock \showarticletitle{The role of replications in empirical software
  engineering}.
\newblock \bibinfo{journal}{{\em Empirical software engineering\/}}
  \bibinfo{volume}{13}, \bibinfo{number}{2} (\bibinfo{year}{2008}),
  \bibinfo{pages}{211--218}.
\newblock


\bibitem[\protect\citeauthoryear{Stel, Dekker, Tripepi, Zoccali, and
  Jager}{Stel et~al\mbox{.}}{2011a}]%
        {stel2011kaplan}
\bibfield{author}{\bibinfo{person}{Vianda~S Stel}, \bibinfo{person}{Friedo~W
  Dekker}, \bibinfo{person}{Giovanni Tripepi}, \bibinfo{person}{Carmine
  Zoccali}, {and} \bibinfo{person}{Kitty~J Jager}.}
  \bibinfo{year}{2011}\natexlab{a}.
\newblock \showarticletitle{Survival analysis I: the Kaplan-Meier method}.
\newblock \bibinfo{journal}{{\em Nephron Clinical Practice\/}}
  \bibinfo{volume}{119}, \bibinfo{number}{1} (\bibinfo{year}{2011}),
  \bibinfo{pages}{c83--c88}.
\newblock


\bibitem[\protect\citeauthoryear{Stel, Dekker, Tripepi, Zoccali, and
  Jager}{Stel et~al\mbox{.}}{2011b}]%
        {stel2011cox}
\bibfield{author}{\bibinfo{person}{Vianda~S Stel}, \bibinfo{person}{Friedo~W
  Dekker}, \bibinfo{person}{Giovanni Tripepi}, \bibinfo{person}{Carmine
  Zoccali}, {and} \bibinfo{person}{Kitty~J Jager}.}
  \bibinfo{year}{2011}\natexlab{b}.
\newblock \showarticletitle{Survival analysis II: Cox regression}.
\newblock \bibinfo{journal}{{\em Nephron Clinical Practice\/}}
  \bibinfo{volume}{119}, \bibinfo{number}{3} (\bibinfo{year}{2011}),
  \bibinfo{pages}{c255--c260}.
\newblock


\bibitem[\protect\citeauthoryear{Xia, Fu, Shu, and Menzies}{Xia
  et~al\mbox{.}}{2020}]%
        {xia2020predicting}
\bibfield{author}{\bibinfo{person}{Tianpei Xia}, \bibinfo{person}{Wei Fu},
  \bibinfo{person}{Rui Shu}, {and} \bibinfo{person}{Tim Menzies}.}
  \bibinfo{year}{2020}\natexlab{}.
\newblock \showarticletitle{Predicting project health for open source projects
  (using the DECART hyperparameter optimizer)}.
\newblock \bibinfo{journal}{{\em arXiv preprint arXiv:2006.07240\/}}
  (\bibinfo{year}{2020}).
\newblock


\end{thebibliography}

\end{document}